# Current Views on Mechanisms of the FLASH Effect in Cancer Radiotherapy

Yuqi Ma[1], Ziming Zhao[1], Wenkang Zhang[1], Jianfeng Lv[1], Junyi Chen[1], Xueqin Yan[1], XiaoJi Lin[2], Junlong Zhang[3], Bingwu Wang[3], Song Gao[3,4], Jie Xiao[5]*, Gen Yang[1,6]*

[1] State Key Laboratory of Nuclear Physics and Technology, School of Physics, Peking University, Beijing 100871, P. R. China
[2] Oncology Discipline Group, the Second Affiliated Hospital of Wenzhou Medical University, Wenzhou 325003, P. R. China
[3] Beijing National Laboratory of Molecular Science, College of Chemistry and Molecular Engineering, Peking University, Beijing 100871, P. R. China
[4] Guangdong Basic Research Center of Excellence for Functional Molecular Engineering, School of Chemistry, Sun Yat-sen University, Guangzhou 510275, P. R. China.
[5] KIRI Precision Particle Therapy Flash Technologies Research Center, 510700, Guangzhou, P. R. China
[6] Medical Cancer Center, The University of Hong Kong-Shenzhen Hospital, Shenzhen 518053, P. R. China

*Corresponding author
Correspondence to: Jie Xiao (15110291500@163.com) and Gen Yang (gen.yang@pku.edu.cn);

## Abstract

FLASH radiotherapy (FLASH-RT) is a new modality of radiotherapy by delivering doses with ultra-high dose rates. FLASH-RT has the ability to suppress tumor growth while sparing normal tissues, known as the FLASH effect. Although FLASH effect has proved valid in various models by different ionizing radiations, the exact underlying mechanism is still unclear. This article summarizes mainstream hypotheses of FLASH effect at physicochemical and biological levels, including oxygen depletion and free radical reactions, nuclear and mitochondria damage, as well as immune response. These hypotheses contribute reasonable explanations to the FLASH effect, and are interconnected according to the chronological order of the organism's response to ionizing radiation. By collating the existing consensus, evidence, and hypotheses, this article provides a comprehensive overview of potential mechanisms of FLASH effect and practical guidance for future investigation in the field of FLASH-RT.

**Keywords**：ultra-high dose rate irradiation, FLASH effect, radiotherapy, mechanism

# Introduction

Cancer is one of the leading causes of human mortality at present[1]. Nowadays, approximately 50% of cancer patients need radiotherapy as a primary or adjunctive therapy[2]. The goal of improving radiotherapy efficacy is to effectively kill tumor tissue while minimizing the damage to peripheral healthy tissue. FLASH radiotherapy (FLASH-RT) with ultra-high dose rate (UHDR, typically ≥40 Gy/s) is an emerging technique to achieve this goal[3, 4].

Exploration into the biological effects of UHDR radiation dates back to the last century. In 1958, Kirby-Smith and Dolphin first demonstrated the decreased chromosomal aberrations in *Tradescantia* microspores and pollen at high dose rates (1 Gy/s and $4.0 \times 10^6$ Gy/s) irradiation compared to low dose rates (≤0.01 Gy/s) irradiation[5]. In 1959, Dewey and Boag found that under hypoxia conditions, the survival fraction of bacteria (*Serratia marcescens*) exposed to UHDR irradiation was higher than that of conventional dose rate irradiation[6]. It was the first report that FLASH irradiation (FLASH-IR) has a protective effect on living organisms. Later, this protective effect of FLASH-IR was demonstrated in mammalian cells[7]. In 2014, a groundbreaking *in vivo* study performed by Favaudon *et al.* revealed that compared with conventional irradiation (CONV-IR, 0.03 Gy/s), FLASH-IR (>40 Gy/s) minimizes radiation-induced lung fibrosis of mice while retaining the toxicity to the lung tumor[4], which is called as FLASH effect now. This work launched a blast of upsurge of physicochemical and biological research in the field of FLASH-RT. Numerous studies on FLASH effect have been carried out, including the mechanism exploration and the clinical trials (Figure 1).

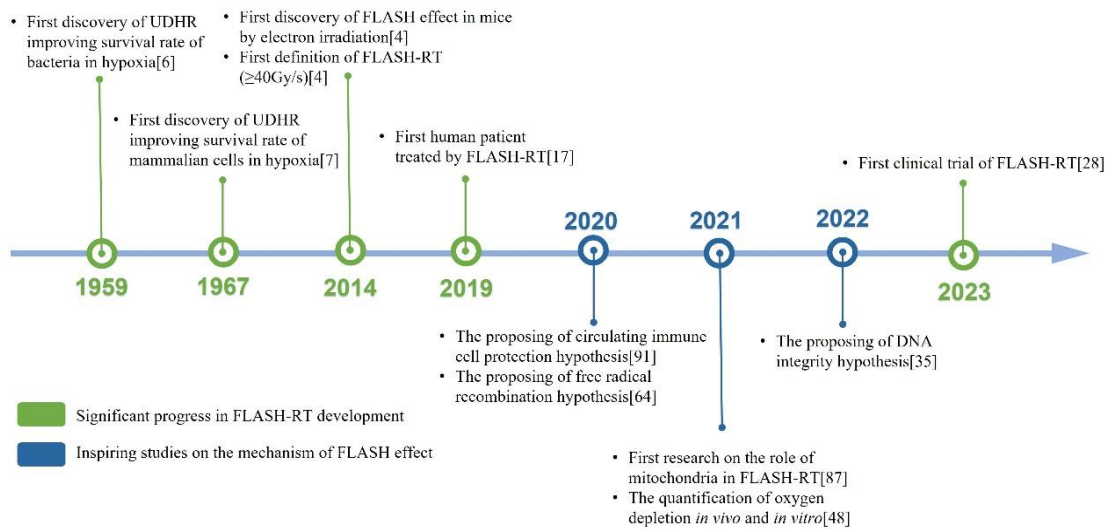

Figure 1 Big events in the development of FLASH-RT.

Based on the ionizing radiation source, UHDR delivery methods employed in FLASH-RT experiments can be classified into three types: electron, proton (as well as heavy ion), and photon (X-ray). FLASH-IR experiments with electron irradiation were performed in the early years[5, 7], and the discovery of FLASH effect[4] stimulated a great number of animal experiments with electron FLASH-IR. The sparing effect of electron FLASH-IR has been demonstrated on a variety of normal tissues[8-14]. Meanwhile, experiments have shown the equivalent tumor control efficacy of electron FLASH-IR compared to CONV-IR in various tumor models[4, 15, 16]. Electron irradiation is also the first type used for FLASH-RT on a human patient. The patient was presented with a multiresistant $CD30^+$ T-cell cutaneous lymphoma and the electron FLASH-RT treatment yielded favorable outcomes[17].

Compared with the electron FLASH, the development of proton FLASH is relatively late. It is noted that as proton beam has high linear energy transfer (LET) and the characteristic of Bragg Peak in tissues, the combination of FLASH-IR and proton irradiation may better benefit normal tissue protection. Since the first demonstration of FLASH effect using proton irradiation[18], proton FLASH has been proved effective on various models[18-24]. Besides, some studies confirmed the effectiveness of proton FLASH-IR utilizing Spread-Out Bragg Peak (SOBP)[19-21]. On the other hand, many experiments unveiled the equivalent[24, 25] or even improved[26, 27] tumor suppression by using proton FLASH-IR compared to CONV-IR in different tumor models. Cincinnati Children's Proton Therapy Center conducted the first clinical trial of proton FLASH on 10 patients with bone metastases (FAST−01)[28]. In this study, the therapeutic effect of FLASH-RT is as efficient as that of CONV-RT, which indicates the clinical feasibility of proton FLASH-RT. As for heavy ion irradiation, Tinganelli *et al.* performed *in vitro* and *in vivo* FLASH studies with carbon ions[29, 30] and revealed the induction of FLASH effect by carbon. Besides, helium ion FLASH-IR was also shown valid in non-tumor cell sparing under hypoxic conditions[31].

X-ray is the most commonly used type of radiation in clinical radiotherapy, and Montay-Gruel *et al.*[32] first demonstrated the X-ray-triggered FlASH effect in mouse brain by using the kilovoltage (kV) X-rays generated at European Synchrotron Research Facilities. However, kV X-rays are considered unfeasible for treating deep-seated tumors because of the rapid decrease of mean dose rate with the penetration depth in tissues[33]. Therefore, megavoltage (MV) X-rays were further developed for the research of FLASH-RT. Based on the MV X-ray generated by electrons bremsstrahlung, researchers delved deeper into the biological effect of X-rays FLASH-RT[34-36].

Nevertheless, whether it's electron FLASH[37, 38], proton FLASH[12, 39], or X-ray FLASH[40], negative experimental results against FLASH effect have been observed both *in vitro* and *in vivo*, which demonstrates that the average/instantaneous dose rate is not the only decisive factor in FLASH effect[41]. The complex and unclear condition of FLASH effect underscores the importance of exploring the specific mechanism behind it. A convincing hypothetical mechanism should explain two questions: how dose rate affects radiotherapy efficacy between FLASH-RT and CONV-RT, and what inherent differences lead to the opposite responses in normal and tumor cells. This paper introduces several mainstream explanations of the mechanism of FLASH effect, including oxygen depletion hypothesis, free radical reaction hypothesis, DNA integrity hypothesis, mitochondrial hypothesis, immunological hypothesis, and other possible mechanisms. These hypotheses contribute reasonable explanations for the FLASH effect but on different spatio-temporal scales (Figure 2).

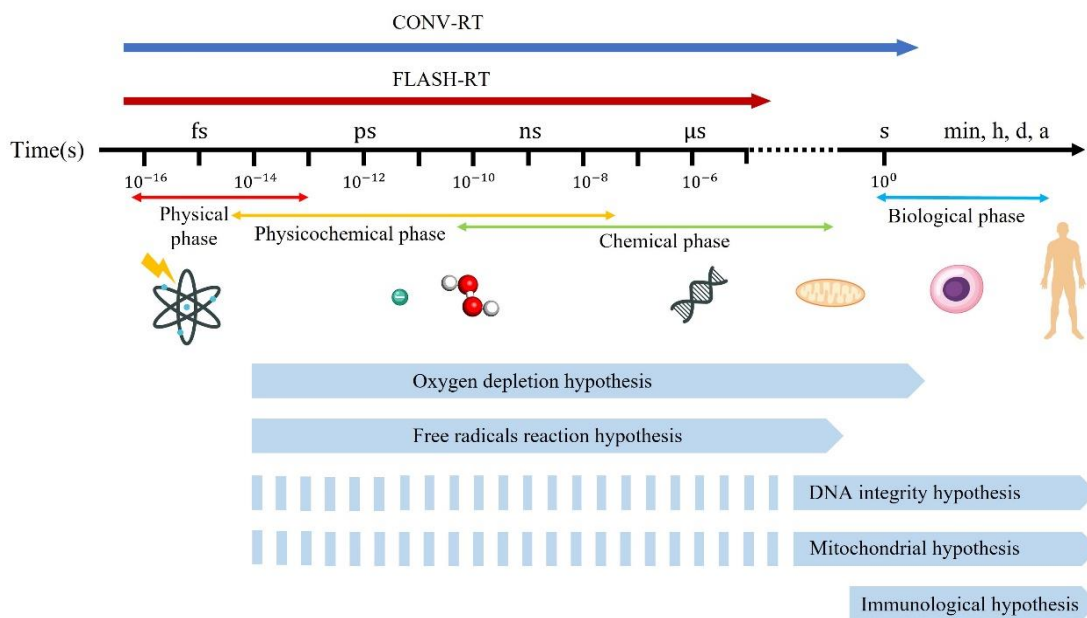

Figure 2 Schematic diagram of the radiation effect at different time scales. The time scale of mainstream hypotheses is shown (the dotted line means not yet proven by experiment or simulation).

## Possible mechanisms of FLASH effect

### Oxygen Depletion hypothesis

The mechanism of oxygen depletion has gained the most popularity among various possible mechanisms. Early during the 1980s, scientists reached a consensus regarding the protective role of low oxygen concentrations during low dose rate irradiation. The difference in biological lethality between normoxic and hypoxic conditions is known as the oxygen enhancement ratio (OER)[42]. As to the FLASH-IR, Dewey and Boag first illustrated the correlation between increasing dose rates and oxygen depletion by comparing the survival curves of bacteria irradiated at UHDR and conventional dose rates under hypoxic conditions[6]. Earlier experiments have shown the relationship between hypoxia and the protective effect of FLASH-IR[7, 43]. Montay-Gruel *et al.*[8] demonstrated that oxygen concentration is a key factor in FLASH effect for the first time in animal experiments. Adrian *et al.*[15] also reported that hypoxia within cancer cells (prostate cancer cell line DU145) contributes to the radio-resistance to FLASH-IR.

Based on these experimental phenomena, researchers hypothesized that the sparing effect of FLASH-RT may result from the rapid depletion of oxygen[44]. Ionizing radiation causes indirect damage by generating free radicals through water radiolysis and following reactions, especially in low LET types of radiation (e.g. X-ray)[45]. By reaction with oxygen molecules, these radicals can form oxygen peroxyl radical (DNA-OO•), which is a more severe and irreparable state of damage[44]. As FLASH-IR consumes oxygen in a short time through the following two reactions: $e^-_{solv}+O_2\rightarrow O_2^-\bullet$, $H\bullet+O_2\rightarrow HO_2\bullet$, it may produce a significant reduction in oxygen concentration, thereby preserving DNA from transitioning into an irreversibly damaged state[46, 47]. On the other hand, CONV-IR delivers doses in a timescale of minutes, which is much longer than the reoxygenation time in cells, resulting in only a minor reduction in oxygen concentration[48]. The unchanged tumor-killing efficacy may be explained by the relatively hypoxic environment in tumors, where the reduction in cell oxygenation after FLASH-IR is comparatively minimal, leading to inadequate changes in radiosensitivity[3] (Figure 3).

The essence of validating oxygen depletion hypothesis is to clarify how scarce should oxygen be to manifest significant protection in normal cells against tumor cells, and estimate that value in clinical practice. Therefore, real-time monitoring of oxygen concentration is imperative, which should ideally have a temporal resolution in milliseconds due to the nature of rapid oxygen depletion during FLASH-IR[49]. Quenching of fluorescent or phosphorescent dyes by oxygen is a common technique for measuring oxygen consumption during FLASH-IR[50]. Cao *et al.*[48] measured and compared the oxygen consumption using Oxyphor 2P probe under electron FLASH-IR and CONV-IR. *In vitro* experiments[48] showed comparable oxygen depletion under FLASH-IR (0.16 to 0.17 mmHg/Gy) and CONV-IR (0.19 to 0.21 mmHg/Gy), which is consistent with previous experimental results[49, 51]. As for *in vivo* experiments[48], the total decrease in oxygen after a single fraction of 20 Gy FLASH-IR was $2.3 \pm 0.3$ mmHg in normal tissue and $1.0 \pm 0.2$ mmHg in tumor tissue, whereas no decrease in oxygen was observed in CONV-IR mode. This study demonstrated that FLASH-IR does deplete oxygen in tissues compared to CONV-IR, but not enough to induce sufficient hypoxia and consequent significant reduction in radiosensitivity, since the physiological oxygen level in tissues is about 38 mmHg[52]. However, it should be noted that in the experiment performed by Cao *et al.*, the temporal resolution of measurement (~150 ms) was much larger than the reaction time of FLASH-IR (~74 ms in the research), and the measurement existed "blind" time, so the actual oxygen depletion may be greater than the measured value[48].

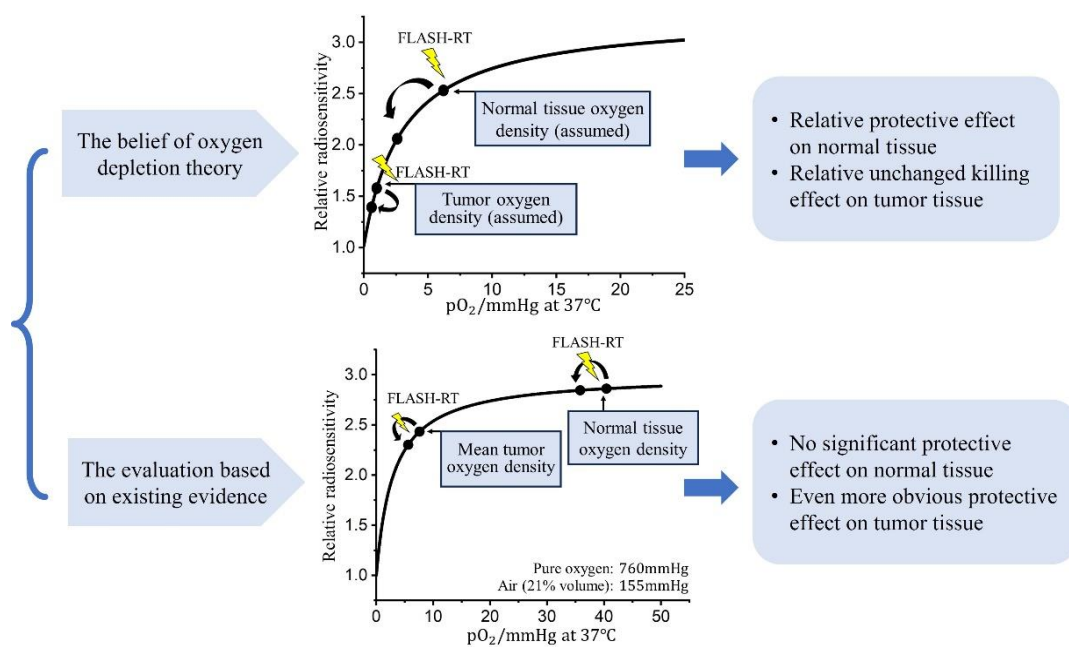

Figure 3 The illustration that oxygen depletion theory cannot fully explain the FLASH effect.

Compared to experimental measurements, simulations offer a more detailed insight into the effects of reactions and parameters, aiding researchers in comprehending the process of oxygen depletion during irradiation. Boscolo et al.[53] did a Monte Carlo simulation focusing on the chemical track evolution of 1 MeV electrons transmitting in water, and the simulated oxygen consumption rate (~0.33 μM/Gy or 0.19 mmHg/Gy) found an agreement with previously measured values[49]. To assess the effect of oxygen consumption on cell radiosensitivity under FLASH-IR, researchers developed relevant models, some of which took into account oxygen diffusion and cell survival curves[46, 54]. Current modeling studies indicate that the typical doses delivered in FLASH experiments (around 20 Gy[4, 8, 13]) are not sufficient to cause adequate oxygen depletion in normal tissues, while only under extreme hypoxia may the effect of oxygen depletion become more pronounced. This corresponds with many *in vitro* studies reporting that only under the condition of relatively low oxygen fraction (below 2% of volume, depending on experiment setup and cell source) there exists a difference between CONV-IR and FLASH-IR[15, 31, 55].

Recent developments have brought forth several challenges to the oxygen depletion hypothesis. The protective effect of FLASH-IR on normal cells has been proved in *in vitro* experiments conducted under normoxia conditions[11, 56, 57], in which oxygen depletion appears to play a minimal role. In addition, since the OER of heavy ions is close to 1, it suggests that the outcomes of heavy ion (e.g. carbon ion[30]) FLASH experiments bear little correlation with oxygen concentration, thus cannot be explained by oxygen consumption. Further, the relationship between radiosensitivity and oxygen concentration is nonlinear, as the physioxia point lies at the low slope of the oxygen concentration-relative radiosensitivity curve while the hypoxia point lies at the high slope[52]. Thus, if the oxygen depletion hypothesis can be explained by the upper portion of Figure 3 (adapted from the schematic diagram presented by Wilson et al.[3]), there should be a significantly great difference in oxygen concentration decrease between normal tissue and tumor tissue post-FLASH-IR. However, by comparing the existing quantification results[48, 53] and the relationship between radiosensitivity and oxygen concentration[52], it suggests that FLASH-IR cannot cause sufficient oxygen depletion in normal tissue, while the radiosensitivity variation in tumor tissue appears to be even more pronounced (Figure 3).

The aforementioned evidence from both theoretical and experimental studies casts doubts on the rationality of oxygen depletion hypothesis. Therefore, we may conclude that FLASH-IR does reduce oxygen concentration in irradiated tissues, but not enough to cause a significant increase in radio-resistance. Taken together, the FLASH effect can only be partially attributed to the oxygen depletion mechanism.

## Free Radicals Reaction Hypothesis

Due to the limitations of the oxygen depletion hypothesis in explaining phenomena such as the FLASH effect on tissue cells in oxygen-rich environments[4], changes in radiosensitivity caused by oxygen concentration variation in cells may not be the primary factor contributing to the FLASH effect. Instead, the impact of radiation on the chemical composition within cells appears to be more intricate and diverse. Previous research has indicated that irradiating water can lead to water radiolysis and following chemical reactions[58]. This provides another avenue for investigating the FLASH effect.

Ionizing radiations with low LET can cause damage to biological molecules through direct energy transfer. Additionally, they can interact with water molecules to generate hydrated electrons ($e^-_{solv}$) and free radicals such as H• and OH•, which then induce cellular damage indirectly. Specifically, the free radicals generated from water radiolysis further react with biological molecules (RH), producing organic free radicals R•. Subsequent reactions of organic free radicals R• with oxygen yield organic peroxyl radicals ROO• (Figure 4), which are widely regarded as significant contributors to cellular damage. Besides, the radiosensitivity associated with oxygen is ultimately believed to be attributed to the action of ROO•[59]. It is worth noting that the $e^-_{solv}$ generated from the interaction in water can react with oxygen to produce $O_2^-$•. By interacting with iron-containing proteins, this species releases unstable active iron ($Fe^{2+}$). $Fe^{2+}$ can form a complex with oxygen, known as $Fe^{2+}$-$O_2$, amplifying oxidative damage through the Fenton reaction. The process converts $H_2O_2$ and ROOH into HO• and RO•, thereby triggering more redox reaction chains[60].

Spitz *et al.*[60] initially explain the different cellular responses to FLASH-RT from the perspective of redox reactions. The key viewpoint is that while the reaction of iron with $O_2^-$• generally comprises only about 1% of other free radical reactions, the concentration of unstable iron in tumor cells is two to four times higher than that in normal cells[61]. This difference leads to increased damage in tumor cells during FLASH-RT compared to normal cells, which can promptly clear $Fe^{2+}$ and ROOH. However, while the theory of Spitz *et al.* elucidates the differences between normal tissue cells and cancer cells under FLASH-RT, it does not account for the protective effect of FLASH-RT on normal tissue cells compared to CONV-RT. Furthermore, there are also some controversies regarding the discussion on the levels of hydrogen peroxide produced by cells under CONV-RT and FLASH-RT in their paper[62].

ROO•, as mentioned earlier, is considered a significant factor in causing cellular damage. On the one hand, it interacts with DNA, inducing chromosomal breaks, aneuploidy, mutations, and thus cell death, while also reacts with unsaturated lipids to generate ROOH, resulting in oxidative damage[63]. On the other hand, R• and ROO• can undergo self-recombination reactions, terminating the aforementioned reaction chain (Figure 4). Building upon redox reactions, Labarbe *et al.* innovatively proposed the free radical recombination hypothesis for FLASH effect[64]. In this framework, recombination reactions and reactions causing cellular damage are considered competitive reactions. The FLASH effect can be explained by UHDR irradiation leading to rapidly elevated concentrations of R• and ROO•, which increase the proportion of recombination reactions and subsequently reduce cell damage. Labarbe *et al.* employed a large number of free radical reaction equations to construct a system of ordinary differential equations to mathematically simulate this hypothesis and quantitatively describe the damage using AUC(ROO•). Although results consistent with the FLASH effect were observed, the hypothesis neglected to explain the differences under FLASH-RT between normal cells and cancer cells.

For the interactions of free radicals, the role of antioxidants in these reactions has not received much attention. Hu *et al.*[65] expanded upon the free radical recombination hypothesis and emphasized the significance of antioxidants, thereby explaining the differences between normal tissue cells and cancer cells under FLASH-RT. Intracellular antioxidants (such as GSH) react with organic free radicals R• to reduce them back to RH, which further decomposes under irradiation to form reaction chains generating a large amount of ROO•. However, ROO• also reacts with a series of antioxidants (such as vitamin E, ascorbate, and glutathione) to produce ROOH[66] (Figure 4). It's evident that the reaction between peroxyl radicals and antioxidants competes with the peroxyl radical recombination reaction, providing a possible explanation for the differences under FLASH-RT between cancer cells and normal tissue cells. Previous experiments have demonstrated that the proportion of the antioxidant GSH in cancer cells is higher than that in normal tissue cells[67]. The alterations in the recombination reactions of free radicals caused by FLASH-RT in cancer cells relative to CONV-RT are minimal, much smaller than those in normal cells. Consequently, FLASH-RT exhibits a cytotoxic effect on cancer cells while conferring protective benefits on normal tissue cells[65].

It's noteworthy that GSH has been shown to mitigate oxidative stress induced by oxidants like hydrogen peroxide. Spitz *et al.*[60] did not explicitly suggest that the process in their theory could potentially lead to ferroptosis, which is characterized by iron-dependent cell death resulting from elevated levels of lipid peroxides catalyzed by the Fenton reaction within cells. GPX4, a GSH-dependent peroxidase, is pivotal in counteracting lipid oxidation[68] (Figure 4). Depletion of GSH may lead to inactivation of GPX4, thereby increasing sensitivity to ferroptosis. Conversely, in Hu's theory of free radicals and antioxidants, GSH reacts with organic peroxyl radicals (ROO·) to generate ROOH, inducing oxidative damage to cells, and acting as a competitive reaction to inhibit radical recombination. In these two theories, GSH appears to play different roles in cell damage or death. Further research and exploration are needed to elucidate the relationship between them and the role of GSH in the FLASH effect.

The key reactions outlined in the aforementioned theories regarding ROS are illustrated in Figure 4. Although these theories seem sound, all the supporting evidences are merely based on mathematical simulations. The work of Labarbe *et al.* has established a benchmark for the mathematical simulation of the free radicals reaction hypothesis[64]. The exceedingly short reaction window poses challenges in experimentally validating these free radicals reaction theories. For example, the generation of free radicals such as HO• from water radiolysis occurs approximately within the range of $10^{-12}$ to $10^{-6}$ seconds after the initialization of irradiation[8]. In addition, due to the complex reaction network of biochemical molecules within cells post ionizing radiations, it is considerably challenging to comprehensively reflect the overall situation of biochemical reactions through the detection of specific components.

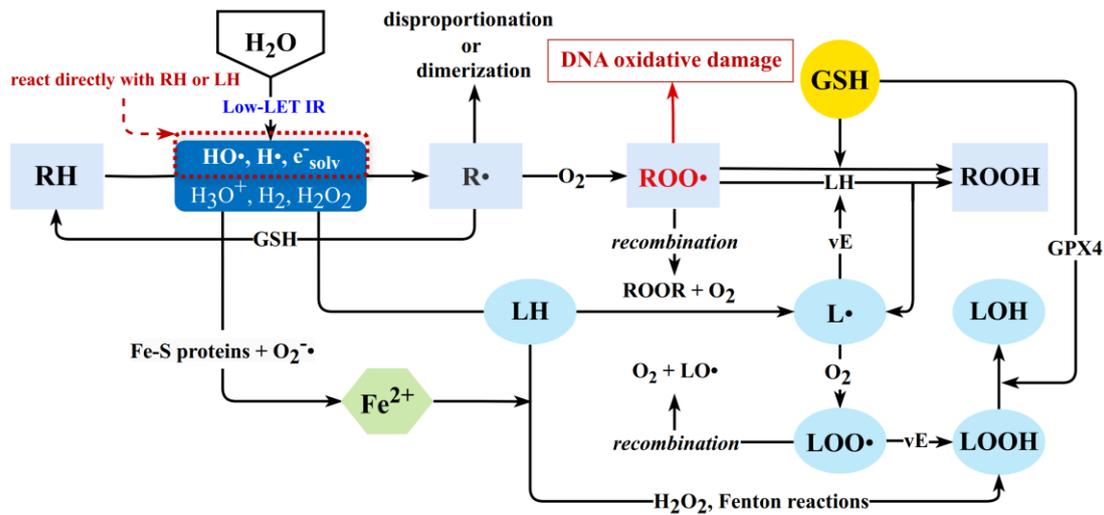

Figure 4 Illustration of free radical reaction caused by irradiation. The products of water radiolysis interact with biomolecules to produce organic free radicals, leading to oxidative damage by organic peroxyl radicals. Hydrogen peroxide may initiate ferroptosis through the Fenton reaction with unstable iron in cells and GSH plays a role in inhibiting ferroptosis. Recombination of radicals or interaction between peroxyl radicals and antioxidants (such as GSH, vitamin E, etc.) can terminate the reaction chain.

## DNA integrity hypothesis

As a result of ROS interaction and chromosome breakage, DNA fragments that leak and accumulate in the cytoplasm trigger the downstream inflammatory process. Through messenger cyclic GMP-AMP synthase (cGAS) and stimulator of interferon genes (STING) to the final secreta type I interferon (IFN-I), this cGAS-STING pathway recruits $CD8^+$T cells to cause immune response and inflammation. Therefore, DNA destruction evokes immune response and may become an explanation for the differences in CONV-RT and FLASH-RT contexts.

Shi et al.[35] disposed the intestinal crypts of mice with 110-120 Gy/s X-ray FLASH, and observed a significant reduction in cytoplasmic double-stranded DNA (dsDNA) levels along with decreased activation of the cGAS-STING pathway compared to 0.03 Gy/s CONV-IR in mouse intestinal crypts. Based on this phenomenon, they proposed the DNA integrity hypothesis, suggesting that the rapid dose deposition of FLASH-IR (~ 100 ms) minimizes the probability of DNA breakage, maintains genomic stability, and reduces cGAS-STING signaling pathway activation. This conclusion corresponds with several recent *in vitro* experiments. Perstin et al.[69] used 16 MeV electron FLASH-IR to irradiate DNA plasmids in water and found a significant DNA double-strand break (DSB) decrease at approximately 46%, which provides compelling *in vitro* evidence for the protective effect of FLASH-IR.

Immunofluorescence analysis of DNA damage response (DDR) proteins is a common method for DNA damage quantification, with γH2AX and 53BP1 widely used as markers for evaluating DSB damage foci[70]. It is essential to consider the timing of peak foci formation (typically 15-30 minutes post-irradiation[71]) when evaluating the extent of irradiation-induced

DSB damage, which is unnoticed by some research. Several studies have reported no significant difference in the induction of γH2AX/53BP1 foci at low doses between FLASH-IR and CONV-IR[35, 57, 72]. However, in the research of Buonanno *et al.*[57], γH2AX generation is reduced with FLASH-IR (1000 Gy/s) at higher doses (20 Gy). The observed variance in foci count between FLASH-IR and CONV-IR may be attributed to FLASH-IR potentially generating clustered DSBs, as proposed by Atkinson *et al.*[73]. It can be inferred that FLASH-IR does not induce fewer DSBs compared to CONV-IR, but rather leads to a higher degree of DSB clustering with more DSBs per focus, resulting in fewer foci. At low doses, the extent of DSB clustering induced by FLASH-IR may not be significant enough to discern a notable difference in foci count. While at higher doses, FLASH-IR may induce more pronounced DSB clustering, leading to a significant reduction in foci count compared to CONV-IR. In addition, Kim *et al.*[16] revealed that FLASH-IR promotes expedited foci repair compared to CONV-IR. Since clustered DSBs have proved more beneficial for efficient repair[74], this phenomenon may be attributed to FLASH-IR inducing a higher degree of DSB clustering. At present, the research on the effect of FLASH-IR on DNA damage mainly focuses on foci quantification, while the quality of DNA damage and repair under FLASH-IR, such as the specific damage pattern and damage repair pathway, still requires further exploration.

Besides explaining how radio-resistance in normal cells be enhanced by FLASH-IR, it's equally important to account for the same killing effect in tumor cells. This may be clarified by the unstable genome and weak DNA damage response (DDR) process of tumor cells[75], which prevents them from repairing conventional damage. Key signaling genes within the DDR pathway are often mutated post-cancerization, including PARP-1 which participates in DNA single-strand break (SSB) repair[76], ATM which delivers ROS signals for radioresistant transitions, and ATR which guarantees regular chromosome segregation through checkpoint responses[77]. Also, the relatively harmless SSBs for normal cells are likely to accumulate in tumor cells, and finally convert to DSBs in the next DNA replication period. Genetic defects render tumor cells vulnerable, lead to considerable DNA fragment leakage under either CONV-IR or FLASH-IR, and subsequently stimulate IFN-I pathway to induce inflammation. Therefore, in the context of tumor cells, the distinction between SSBs and DSBs may be less relevant, with dose emerging as the primary determinant of response, rather than dose rate[78]. Nevertheless, a comprehensive understanding of the distinct responses in normal versus tumor cells under FLASH-IR requires further elucidation of the specific damage patterns and repair pathways engendered by FLASH-IR.

In addition to nuclear DNA, mitochondrial DNA may also play an important role in the FLASH effect, which will be discussed in the next section of this article.

## Mitochondrial Hypothesis

Mitochondria play pivotal roles in a variety of cell death pathways. First of all, mitochondria serve as the key organelles of cell endogenous apoptosis. Upon exposure to ionizing radiation, mitochondria generate excess reactive oxygen species (ROS), leading to mitochondrial outer membrane permeabilization (MOMP) mediated by pro-apoptotic Bcl-2 proteins BAX and BAK[79]. Consequently, mitochondrial proteins such as cytochrome c (cyt

c) are released into the cytoplasm, where cyt c binds with APAF-1 to recruit caspase-9 to form apoptosome, inducing apoptotic reaction[80]. Apoptosis is generally considered to be an active programmed process of autonomous cellular dismantling which avoids eliciting inflammation[81]. In addition to apoptosis, mitochondria are also involved in multiple pro-inflammatory response pathways, which are mainly associated with passive and accidental cell death. One noteworthy pathway is that MOMP can promote the release of mitochondrial DNA (mtDNA) into cytoplasm, which evokes the cGAS-STING pathway and induces the release of IFN-I, leading to downstream inflammatory responses[82]. It has been shown that apoptosis-related caspases inhibit the induction of IFN-I by mtDNA[83], suggesting a role for caspase-mediated apoptosis in preventing dying cells from triggering a host immune response[84]. This interplay between caspase-mediated apoptosis and IFN-I-mediated inflammation has also been demonstrated in ionizing radiation studies, in which the blocking of caspase-mediated apoptosis results in the increased release of mtDNA, the enhancement of immunity[85], and the induction of abscopal responses to radiotherapy[86]. These findings offer insights into strategies for enhancing the anti-tumor immune response or attenuating the normal tissue inflammatory response caused by radiotherapy.

Because of the vital role of mitochondria in regulating cell death and immune response, researchers are curious about their potential role in FLASH effect. Han et al.[87] conducted pioneering research on this topic, employing FLASH-IR ($>10^9$ Gy/s ultrafast laser-generated particles) and CONV-IR (0.05 Gy/s $Co_{60}$ γ radiation) to irradiate normal (cyt $c^{+/+}$) and cyt c-defective (cyt $c^{-/-}$) mouse embryonic fibroblast cells. Compared with normal (cyt $c^{+/+}$) cells, the proportion of late apoptosis and necrosis in cyt c-deficient (cyt $c^{-/-}$) cells was significantly lower under FLASH-IR in both hypoxia and normoxia, which implied that the loss of mitochondrial function may increase the resistance of cells to FLASH-IR. Guo et al.[56] carried out a comprehensive study that irradiated normal lung fibroblasts (IMR90) with protons at the condition of FLASH (100 Gy/s) or CONV (0.33 Gy/s) under ambient oxygen concentration (21%). Compared with CONV-IR, FLASH-IR improved cell survival and prevented mitochondria damage, while conversely the cell viability and mitochondrial morphology of lung cancer cells (A549) would be negatively affected by both FLASH-IR and CONV-IR. The examination of the fate of irradiated cells showed that FLASH-IR did lead to apoptosis and possibly autophagy, while CONV-IR mainly induced necrotic cell death. Based on the results, Guo et al. proposed a possible mechanism that proton FLASH-IR induces p-Drp1 to preserve normal cellular mitochondrial function, while CONV-IR leads to dephosphorylation of p-Drp1. The presence of Drp1 in mitochondria leads to mitochondrial fission and cell necrosis.

To further explore the role of mitochondria in FLASH effect, Lv et al.[88] conducted an inspiring study unveiling the regulation of mitochondria-mediated apoptosis and inflammatory pathways by FLASH-IR. They observed that compared to the low dose rate electron irradiation (0.36 Gy/s), FLASH-IR (61 or 610 Gy/s) amplifies cyt c leakage from mitochondria in human breast cells MCF-10A, which elicits substantial caspase activation but suppresses both the cytosolic mtDNA accumulation and IFN-β production. This result indicates that FLASH-IR may enhance the programmed apoptosis mediated by the cyt c-caspases chain and inhibit the IFN-I inflammatory response mediated by the mtDNA-induced cGAS-STING pathway, thus alleviating the immune damage of normal tissues post-irradiation (Figure 5). In contrast, the cyt c leakage in carcinoma cells MDA-MB-231 after electron irradiation is limited, especially for the case of FLASH-IR, resulting in less cytosolic cyt c but stronger cGAS-STING activation than those in MCF-10A cells. Lv *et al.*[88] explained this difference in cyt c leakage by the Warburg effect[89] that cancer cells tendentiously produce energy by aerobic glycolysis in cytoplasm rather than oxidative phosphorylation and citric acid cycle in mitochondria. Interestingly, in their research MCF-10A cells show no significant change in mitochondrial morphology post-irradiation, and even enhanced mitochondrial fission is observed in MDA-MB-231 cells post-FLASH-IR. Furthermore, the variation of cyt c leakage with increased irradiation dose rate was found to be inconsistent with the mitochondrial morphology change. These phenomena cannot be explained by the hypothesis about Drp1 pathway proposed by Guo *et al*[56]. Based on these findings, Lv *et al.*[88] proposed the hypothesis of electron transport chain (ETC) disruption, that is FLASH-RT stimulates extensive cascade feedback of ETC dysfunction and cyt c detachment from cardiolipin in the normal cells, which enhances the cyt c leakage from mitochondria to cytosol. To prove the correctness of this hypothesis, the function of ETC or the change of mitochondrial membrane potential (MMP) post-irradiation needs to be further investigated.

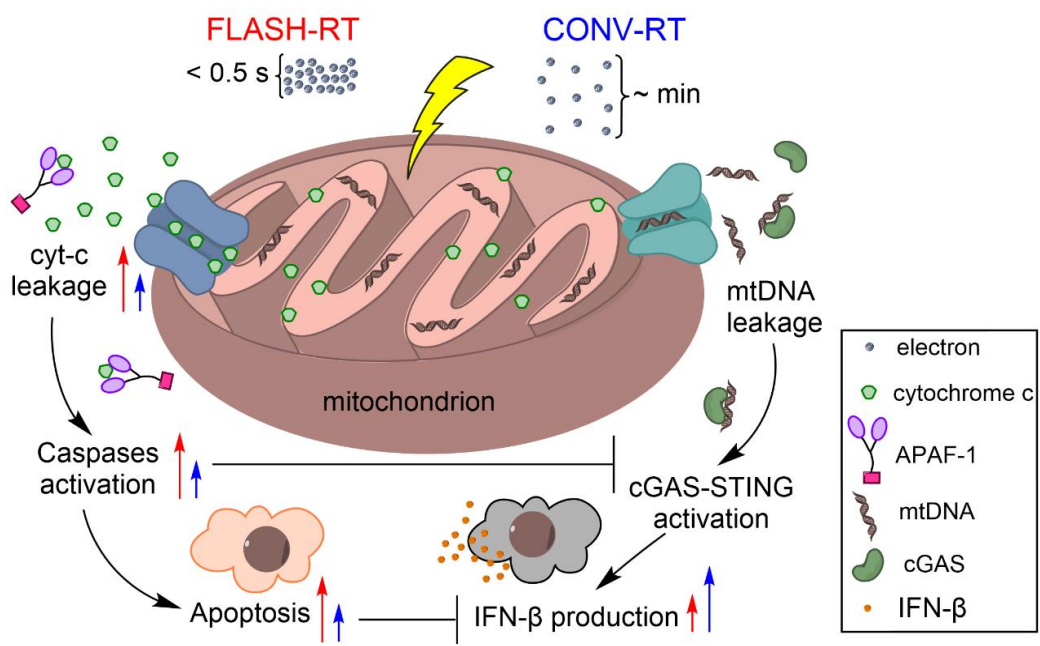

Figure 5 Illustration of the mitochondrial hypothesis[88]. FLASH-RT regulates the mitochondria-mediated apoptosis and inflammatory pathways to cause FLASH effect.

An alternative perspective on the potential function of mitochondria in FLASH effect is from the perspective of mitochondrial ROS (mtROS)[90]. Researchers believe that FLASH-IR can reduce cell damage by limiting the increase of mtROS compared with CONV-IR in normal cells. As for tumor cells, the overproduction of mtROS renders them more susceptible to the increase of mtROS level. The specific mechanism here can be referred to in the section of free radicals reaction hypothesis.

## Immunological Hypothesis

Jin *et al.*[91] proposed the circulating immune cell protection hypothesis, highlighting the strong sparing effect of FLASH-RT on circulating immune cells through modeling and computation. They found that the number of lymphocytes affected by FLASH-RT in blood circulation is significantly lower than that by CONV-RT due to the relatively short irradiation time of FLASH-RT. Computation revealed a reduction in the killing rate of circulating immune cells from 90%-100% under CONV-RT to 5-10% under FLASH-RT. The sparing effect of circulating immune cells diminishes the destruction of the body's immune system, helps the repair of damaged tissue cells, and thus reduces the level of tissue damage. To validate this theory, Galts *et al.*[92] exploited a dosimetric blood flow model to simulate the dose of circulating lymphocytes during CONV and FlASH pencil-beam scanning-based (PBS) intensity modulated proton therapy (IMPT) for brain tumors. The favorable simulation results verified the circulating immune cell protection by proton FLASH-RT during intracranial therapy.

To validate the computation and simulation results, experimental observation of the protective effect of FLASH-RT on circulating immune cells is imperative. Venkatesulu *et al.*[38] regularly measured the number of peripheral blood lymphocytes after irradiating the heart and spleen of mice with CONV (0.1 Gy/s) and FLASH (35 Gy/s) electron beams. Contrary to expectations, the FLASH-RT treated group exhibited even greater lymphocyte decline and higher toxicity to normal tissues, showing no sparing effect. Zhang *et al.*[39] irradiated part of the abdomen of mice utilizing 120 Gy/s proton FLASH-RT, but no normal tissue sparing or depletion of circulating lymphocytes was observed. These negative results challenge the immune system protection hypothesis, suggesting that the effect of FLASH-RT on peripheral blood circulating immune cells may depend not only on the proportion of immune cells irradiated, but also on the complex damaging effects of UHDR irradiation. Interestingly, in experiments yielding negative results for the protective effects of FLASH-RT on circulating immune cells, there were also no observed protective effects of FLASH-RT on normal tissues. Conversely, experiments supporting the FLASH effect did not detect changes in the number of circulating immune cells. Therefore, the validity of this circulating immune cell protection hypothesis needs further validation and exploration.

FLASH-RT has the potential to alter the immune cell composition within irradiated tissue, thereby affecting anti-tumor immunity. Rama *et al.*[26] conducted a FLASH experiment on a model that Lewis Lung Carcinoma (LLC) cells were inoculated into the left lung of mice. They observed the increased recruitment of $CD3^+$ T lymphocytes, particularly $CD4^+$ and $CD8^+$ cells, from peripheral tumor margins to the tumor core under FLASH-IR compared with CONV-IR. Based on the similar LLC mice model, Kim *et al.*[16] and Shukla *et al.*[27] respectively proved that FLASH-IR increases the infiltration of cytotoxic $CD8^+$ T lymphocytes within the tumor in comparison with CONV-IR, thus guarantees a better tumor-killing effect. Furthermore, in the research done by Shukla *et al.*[27], FLASH-IR was more effective in decreasing the percentage of immunosuppressive regulatory T cells (Tregs) in T lymphocytes, decreasing pro-tumorigenic M2-like macrophages, and increasing anti-tumor M1-like macrophages. Some studies have revealed the impact of FLASH-IR on the number of macrophages in neural tissue. They reported that compared with COVN-IR, FLASH-IR reduced the activation of microglia (macrophages in nerve tissue), which in turn reduced the induction of neuroinflammation[14, 21, 93].

FLASH-RT may alter the expression of certain cytokines, thereby contributing to FLASH effect. Transforming growth factor-β (TGF-β) is a multifunctional cytokine which plays an important role in regulating the immune system and tumor growth. The increase in TGF-β after exposure to ionizing radiation is proved to have side effects on different normal tissues, such as fibrosis induction[94]. Studies have shown that compared to CONV-IR, FLASH-IR can reduce TGF-β production in normal tissues, including mouse lung[4], mouse skin[22, 23], canine skin[23], and human lung fibroblasts[57]. This may explain the protective effect of FLASH-IR on normal tissues. In addition, TGF-β exhibits a dual role in tumor regulation. TGF-β plays a tumor-suppressive role in normal and early-stage cancer cells mainly by promoting apoptosis and inhibiting cell cycle progression, while it also plays a tumor-promoting role in late-stage cancer cells mainly by inducing proliferation, invasion, angiogenesis, metastasis, and immune suppression[90, 94]. Investigating changes in TGF-β levels in tumor tissues post-FLASH-IR could provide insights into its role in maintaining lethality in tumor cells. In addition to TGF-β, FLASH-IR was also proved to induce variations in other cytokines compared with CONV-IR, such as Cxcl-1, G-CSF, GM-CSF, IL-1β, IL-4, IL-6, IL-10, TNF-α, and KC/GRO[22, 36, 93]. Various cytokines influence the immunological effects of FLASH-IR through a complex interaction network. The specific impact details need to be further investigated.

In recent years, a kind of unique clinical radiotherapy response, abscopal effect, has attracted much attention, which refers to the tumor regression at sites distant from the irradiated volume in radiotherapy[95]. The principle of abscopal effect is thought to be related to immunogenic cell death (ICD), mediated through two main mechanisms[96]. The first is the release of tumor associated antigen (TAA) and tumor specific antigen (TSA). The second is the release of specific damage associated molecular patterns (DAMPs), such as calreticulin exposure on the surface of cell membranes, as well as the extracellular release of heat shock proteins, high mobility group protein B1, and adenosine triphosphate[97]. While there's substantial knowledge about ICD generation in CONV-RT, limited research has focused on ICD under FLASH-RT conditions and the subsequent abscopal effect. Shi *et al.*[35] demonstrated that with the assistance of anti-PD-L1 therapy, X-ray FLASH-IR can produce the same inhibitory effect on both primary and secondary tumors as CONV-IR. This implies that FLASH-RT also has the potential to induce abscopal effect. However, further verification of abscopal effect under the condition of FLASH-RT necessitates experimental evaluation of various immune cells and immune molecules such as TAA, TSA, and DAMPS involved in ICD.

### Other possible mechanisms

The effect of FLASH-IR on stem cells (both normal tissue stem cells and tumor stem cells) may also be an important part of the FLASH effect. Montay *et al.*[9] first supposed that FLASH-RT might protect neural stem cells based on the known sparing effect of FLASH-IR on neural tissues. Later, Vozenin *et al.*[13] demonstrated the sparing effect of FLASH-IR on porcine skin epidermal stem cells. Stem cells typically reside in hypoxic niches, distant from vasculature, leading to relatively radio-resistance. Pratx et al.[46] stated that the FLASH effect might stem from the specific sparing of stem cells through modeling methods, as the typical oxygen tension in hypoxic stem cell niches overlaps with the range where their model predicts significant radioprotection from oxygen depletion associated with FLASH-RT. Therefore, the sparing effect of FLASH-IR on normal tissues may be related to its protective influence on normal tissue stem cells. As for tumor stem cells, previous studies have highlighted their radio-resistance to CONV-IR[98]. Yang et al.[99] first demonstrated that tumor stem cells (MCF-7 cancer stem cells) were more radioresistant than normal tumor cells (MCF-7) under proton FLASH-IR. Their subsequent experiments tracking lysosomes and autophagy revealed higher levels of lysosomes and autophagy in tumor stem cells. Therefore, the heightened radiotolerance of tumor stem cells may be associated with increased lysosomal-mediated autophagy as well as reduced apoptosis, necrosis, and pyroptosis. The radioresistant characteristic of tumor stem cells provides clues for enhancing FLASH effect and optimizing future clinical FLASH therapy.

Several studies have shown the potential of FLASH-IR to spare blood vessels. The first report on FLASH effect revealed its ability to prevent acute apoptosis in blood vessels and bronchi in mice compared to COVN-IR[4]. FLASH-IR was found to outperform CONV-IR in protecting the integrity of brain microvessels, which may partly elucidate its role in preserving cognitive function[10, 21]. The protective effect of FLASH-IR on blood vessels in normal tissues corresponds with its sparing effect on normal tissues. As for blood vessels in tumor microenvironment (TME), Kim et al.[16] found that in the mouse model with LLC tumor transplantation, while CONV-IR induced reversible collapse of blood vessels (detected 6 h & 48 h post-irradiation), FLASH-IR won't cause a rapid or reversible collapse of blood vessels. In 2001, Jain introduced the concept of vascular normalization[100], indicating that the judicious use of antiangiogenic therapies does not block blood vessels, but rather reverts the severely abnormal structure and function of the tumor vasculature to a more normal state. Tumor vascular normalization enhances tumor perfusion, thus increasing oxygen and drug delivery to the tumor and augmenting the effect of radiotherapy and chemotherapy. From this point of view, the states of tumor blood vessels before and after irradiation, whether they exhibit normal or abnormal characteristics, will contribute to the anti-tumor efficacy. Interestingly, in the work done by Kim et al.[16], there were no differences observed in the CD31-positive vessels number, caspase-3 levels, and hypoxia levels between the treatment of FLASH-IR and CONV-IR, indicating that the state of tumor blood vessels remained unconverted. Further investigation into the effect of FLASH-IR on tumor blood vessels is needed.

## Conclusion and prospects

This article discusses the possible mechanisms of FLASH effect from the perspectives of oxygen depletion, free radicals, DNA damage, mitochondria, immunity, and other possible mechanisms. Importantly, the mechanisms proposed above are not independent of each other, but rather interconnected according to the chronological order of the organism's response to ionizing radiation (Figure 2). Upon exposure to ionizing radiation, the process of radiolysis and following interactions produce free radicals, which will cause damage to DNA and result in dsDNA leakage into the cytoplasm. This indirect DNA damage can be enhanced by the presence of oxygen, thus closely associated with oxygen concentration. As for mitochondria, the radioactive process triggers the leakage of both mtDNA and cyt c. The presence of cytosolic dsDNA and cyt c can modulate downstream immune responses via relevant signaling pathways, ultimately influencing the lethality of radiation exposure. We may organize and elucidate the underlying mechanisms of the FLASH effect along this cascade of radiation responses. The instantaneous dose delivery of FLASH-RT prompts a rapid increase in free radical concentration, and diminishes the proportion of peroxyl radicals through free radical recombination. Consequently, this mitigates damage to biomolecules including DNA, and reduces the leakage of dsDNA into the cytoplasm. Meanwhile, within mitochondria, the rapid escalation of free radicals during FLASH-RT promotes the dissociation and leakage of cyt c. The reduction in cytoplasmic dsDNA attenuates downstream inflammatory pathways, whereas the increase of cyt c enhances the apoptotic pathways related to programmed cell death, thus ultimately reducing the tissue toxicity from FLASH-RT. However, in the case of tumor tissues, distinctive characteristics such as different redox levels, DNA stability, or the Warburg effect, may lead to a breakdown in one of the aforementioned mechanisms, thereby preserving the unaltered response to FLASH-RT.

In recent years, the FLASH effect has attracted a lot of attention for its great application potential in clinic. Based on the above analysis and discussion, we summarize three key issues in the FLASH-RT field. The first issue is the exact thresholds for triggering the FLASH effect. At present, determining the beam parameters yielding FLASH effect relies on semi-empirical approaches, and in some cases dose rate does not seem to be the sole indicator, which makes the research and application of FLASH-RT more empirically based. The second issue is the underlying mechanism of FLASH effect. Each hypothesis discussed in this article can reasonably explain the FLASH effect under certain conditions, but not convincing enough for all cases. The third issue is the clinical application of FLASH-RT. It is important to explore its optimal integration into clinical practice and the potential combination with other therapies, such as immunotherapy. Among the above three issues, the most fundamental one is the unclear mechanism of FLASH effect, which is the original intention of writing this article. Both the free radical production and DNA damage at the physicochemical stage, along with the activation of related signaling pathways and immune responses at the biological stage, are worth exploring.

When assessing the feasibility of FLASH-RT, irradiation parameters including average/instantaneous dose rate, total dose, delivery time, beam energy, pulse count, and pulse rate; along with the properties of irradiated models including species/cell line, irradiated regions, irradiation depth, tumor status, and surviving environment, should be comprehensively taken into consideration. In recent years, with the deepening understanding of FLASH effect, *in vitro* experiments aiming to study the mechanisms of FLASH effect have gradually sprung up. When performing *in vitro* experiments, particular attention should be paid to the oxygen concentration, whether it is normoxia, physoxia, or hypoxia[41]. Different oxygen concentration leads to different radiosensitivity, which may differ from the *in vivo* situations. In addition, it's crucial to note that *in vitro* experiments typically exclude the involvement of immune system, which may impact cellular responses to irradiation and should be considered when analyzing results.

Clinical trials of FLASH-RT using both electrons and protons have achieved promising initial success, motivating further exploration into the physicochemical and biological mechanisms of the FLASH effect. An in-depth understanding of these mechanisms will facilitate the improved clinical implementation of FLASH-RT, thereby enhancing the therapeutic efficacy of radiotherapy.

## Acknowledgements

We express indebtedness to anonymous reviewers for their valuable and constructive comments. This study was partially supported by the National Natural Science Foundation of China (grant nos. 12375334), and National Key Research and Development Program of China (grant nos. 2023YFC2413200/2023YFC2413201 and 2019YFF01014402).

## Competing interests

All other authors declare no competing interests.